\documentclass[english,floatfix,twocolumn,superscriptaddress,nofootinbib,showpacs,prd]{revtex4-2}
\usepackage[paperwidth=210mm,paperheight=297mm,centering,hmargin=2cm,vmargin=2cm]{geometry}
\usepackage[utf8]{inputenc}
\usepackage[T1]{fontenc}
\usepackage{lmodern}
\usepackage{amsmath}
\usepackage{amssymb}
\usepackage{graphicx}
\usepackage{subcaption}
\usepackage{esint}
\usepackage{dcolumn}
\usepackage{babel}
\usepackage{xcolor} 
\usepackage{csquotes}
\usepackage{marvosym}
\usepackage{fontawesome5}
\usepackage{hyperref}
\usepackage[overload]{textcase}
\usepackage[symbol,hang,flushmargin]{footmisc}

\DefineFNsymbolsTM{otherfnsymbols}{
\Bat 	\dagger  \Radioactivity	\ddagger  \Lightning	\textparagraph  \Biohazard	\textbardbl	\MineSign \circ  \Laserbeam  \mathparagraph
}

\setfnsymbol{otherfnsymbols}

\newenvironment{seqn}{\equation\aligned}{\endaligned\endequation}

\newcommand{\be}{\begin{seqn}}
\newcommand{\ee}{\end{seqn}}

\newcommand{\bea}{\begin{eqnarray}}
\newcommand{\eea}{\end{eqnarray}}

\newenvironment{arabicfootnotes}
  {\par\edef\savedfootnotenumber{\number\value{footnote}}
   
   \setcounter{footnote}{0}}
  {\par\setcounter{footnote}{\savedfootnotenumber}}

\begin{document}
%%%%%%%%%
%
%
%
%
%%%%%%%%%%%%%%%%%%%%%%%%%%%%%%%%%%%%%%%%%%%%
\title{Linearized Stability of Harada Thin-Shell Wormholes}

\author{Hassan~Alshal}
\email{halshal@scu.edu}
\affiliation{Department of Physics, Santa Clara University, Santa Clara, CL 95053, USA.}

\author{Leyang~Ding}
\email{lding4@scu.edu}
\email{lyding@ucdavis.edu}
\affiliation{Department of Physics, Santa Clara University, Santa Clara, CL 95053, USA.}
\affiliation{Department of Physics and Astronomy, UC Davis, One Shields Avenue, Davis, CA 95616}

\author{Adelina~Hernandez}
\email{ahernandez2@scu.edu}
\affiliation{Department of Physics, Santa Clara University, Santa Clara, CL 95053, USA.}

\author{Leo~A.~Illing}
\email{lilling@scu.edu}
\affiliation{Department of Physics, Santa Clara University, Santa Clara, CL 95053, USA.}

\author{Ivar~Rydstrom}
\email{irydstrom@scu.edu}
\email{irydstrom@berkeley.edu}
\affiliation{Department of Physics, Santa Clara University, Santa Clara, CL 95053, USA.}
\affiliation{Department of Physics, University of California, 366 Physics North MC 7300, Berkeley, CA, 94720-7300}

\begin{abstract}

\begin{center}
\textbf{ABSTRACT}
\end{center}

\par\noindent
Using Darmois-Israel-Sen junction conditions, and with help of Visser's cut-and-paste method, we study the dynamics of thin-shell wormholes that are made of two conformally Killing gravity (a.k.a Harada gravity) black holes.  We check the energy conditions for different values of the new parameter that Harada introduced, as alternative for dark energy. We examine the radial acceleration to reveal the attractive and repulsive characteristics of the thin-shell wormhole throat. We consider the dynamics and stability of the wormhole around the static solutions of the linearized radial perturbations at the wormhole throat. Finally, we determine the regions of stability by applying the concavity test on the ``speed of sound'' as a function in the throat radius and other spacetime parameters, particularly the new Harada parameter.
\end{abstract}
\maketitle
\begin{arabicfootnotes}
%
%
%
%
%%%%%%%%%%%
\section{Introduction}
\par\noindent
Traversable Lorentzian wormholes \cite{Morris:1988cz,Visser:1995cc} are known as solutions for Einstein's field equations. The Conventional way to find a wormhole solution is to select some equations of state of matter such as Chaplygin gas \cite{Lobo:2005vc}, phantom energy \cite{Sushkov:2005kj, Lobo:2005yv}, and/or quintessence \cite{Lobo:2006ue}. Then one can apply geometrical constructions such as thin-shell spacetimes \cite{Poisson:1995sv}, evolving wormholes \cite{Kar:1995ss}, rotating spacetimes \cite{Teo:1998dp}, dust shell wormholes \cite{Lobo:2004rp}, and/or Casimir wormholes \cite{Garattini:2019ivd} in order to alleviate the violation of energy conditions associated with the \textit{flaring-out} condition that distinguishes wormhole solutions from other solutions. A thin-shell is a singular boundary hypersurface with energy-momentum conditions as stabilizers for such surface. If a thin shell is used to connect two separate spacetime with flaring-out condition, this thin shell forms a thin-shell wormhole. There are numerous studies that consider different black holes creating thin-shell wormholes in de-Sitter (dS) and anti-de-Sitter (AdS) spacetimes \cite{Kuhfittig:2010pb, Rahaman:2011yh, Sharif:2013xta, Sharif:2014ria, Eid:2015pja, Ovgun:2017jzt}. Stability of these thin-shell wormholes are examined too \cite{Ishak:2001az, Lobo:2005zu, Eiroa:2007qz, Eiroa:2008ky, Lemos:2008aj, Dias:2010uh, Eiroa:2011nd, Sharif:2013nka, Mazharimousavi:2014gpa, Lobo:2015lbc, Eid:2016axb, Ovgun:2017jip, Amirabi:2017buh, HabibMazharimousavi:2017zlc, Eiroa:2017nar, Tsukamoto:2018lsg, Forghani:2019wgt,Halilsoy:2013iza, Sharif:2016gyb,Alshal:2019owh}.  Also, one should consider several other examples of wormhole solutions that do not require exotic matter such as those supported by the Dirac-Maxwell fields \cite{Konoplya:2021hsm,Churilova:2021tgn} and those supported by applying different geometrical modifications to the usual spacetime \cite{DiGrezia:2017daq,Bronnikov:2018uje,Kuhfittig:2020zmp,DeFalco:2021ksd,Klinkhamer:2022rsj,DeFalco:2023twb}. It is worth noting that the stability of wormholes is an old issue in theories of gravity, and until this moment there is no wormhole solution for which stability has been realized under \emph{general} perturbations, see the examples in Ref. \cite{Gonzalez:2008wd, Bronnikov:2012ch, Cuyubamba:2018jdl} and the cited references there.\\
\par\noindent
Recently, Harada targeted three main obstacles in Einstein's general theory of relativity \cite{Harada:2023rqw}. These obstacles are: i. the nature of the cosmological constant as an integration constant, ii. the derivation of conservation of energy-momentum tensor rather than being as conjectured an ad hoc, and iii. the usage of conformal flat metric as a vacuum solution and the unphysical solutions associated with it. Harada new gravity theory shows that accelerating expansion of the universe naturally appears as a consequence of the beyond general relativity gravitational field equations even in the absence of the conventional cosmological constant and/or dark energy \cite{Harada:2023afu}. Harada endeavors are considered a conformal Killing gravity theory in which the energy-momentum tensor is corrected by a divergenceless conformal Killing tensor, and thus the theory is compared to Cotton gravity \cite{Mantica:2023stl}. Harada theory comes with many physical consequences \cite{Mantica:2024mun,Mantica:2024sdy} such as generalized solutions of Schwarschild-Reissner-Nordström-AdS and regular black holes \cite{Junior:2023ixh,Barnes:2023uru}, non-asymptotically flat traversable wormholes \cite{Clement:2024xmr}, black bounces \cite{Junior:2024vrv}, and pp-waves \cite{Barnes:2024vjq}.
\par\noindent
We are interested in studying the stability of thin-shell wormholes in Harada gravity theory for the following two main reasons. The first one is that finding a stable soltuion to traversable wormholes in conformal Killing gravity is a prominent result because such solution is not allowed by the null energy condition in the context of the conventional general theory of relativity \cite{Morris:1988tu}. The second reason is that there is a class of wormhole solutions of conformal Killing gravity
found in Ref. \cite{Clement:2024xmr} the does not demand the support of exotic matter or dark energy to stabilize the wormholes; and this stimulates considering the thin-shell wormholes within the same theory. In section ($\mathbf{II}$), we use Visser's cut-and-paste technique \cite{Visser:1989kh, Visser:1989kg}, together with Darmois-Israel-(Sen) junction conditions \cite{Sen:1924,Israel:1966rt,Darmois:1927rt,Mansouri:1996ps}, to connect two conformal Killing gravity regions of spacetime through a thin shell. The cut-and-paste method with the junction conditions are comprehensively reviewed in Ref. \cite{2740310}. The methodology of cut-and-paste provides the advantages of the employment of lesser amount of exotic matter, and hence traveling objects through this wormhole avoid the regions occupied by exotic matter \cite{Visser:1989kh}. This results in confining the exotic matter at the thin-shell regions similar to the matter studied in other examples of thin-shell wormholes \cite{Eiroa:2007qz,Sharif:2013tva,Sharif:2016gyb,Godani:2022hik,Eid:2023fbh,Eid:2023wrd,Sharif:2021ugf,Sharif:2021zzr,Sharif:2021cvr,Sharif:2020nhc,Eiroa:2003wp,Bejarano:2016gyv}. Also, we study the components of the hypersurface energy-momentum tensor using the extrinsic curvature components. We utilize these components to obtain the stress and pressure then comment on the violation of energy conditions because of the exotic matter at the wormhole throat. From the discontinuity in the extrinsic curvature, we compare the effect of extra fourth order term in Harada metric on the spacetime to the effect of the usual second order cosmological constant term on spacetimes studied in Ref. \cite{Lobo:2003xd}. Additionally we comment on the physics of attraction and repulsion on the wormhole throat in terms of the acceleration.
\par\noindent
In section ($\mathbf{III}$), we analyze the linearized stability of Harada thin-shell wormhole by implementing the concavity test on the ``speed of sound'' as a function of spacetime parameters: the mass, the cosmological constant, and the new Harada parameter that corresponds to the fourth order term in the metric. Then, we visualize the change in stability regions upon varying the value of Harada parameter while we fix the values both mass and cosmological constant.\\
\par\noindent
In section ($\mathbf{IV}$) we summarize the results of the previous two sections and comment on them. 
%%%%%%%%%%%%
%
%
%
%
%%%%%%%%%%%
\section{Dynamics of Thin-Shell Wormhole}
%%%%%%%%%%%%
%
%
\par \noindent
The Harada black hole can be constructed \cite{Harada:2023rqw} starting with a static spherically symmetric metric
\be\label{eq.1}
ds^2_{\text{Schw.}}=&- e^{\nu (r)}dt^2+e^{-\nu (r)} dr^2\\
&+r^2d\theta^2+r^2 \sin^2(\theta)d\phi^2~.
\ee
The field equations are derived from the Harada totally symmetric tensor
\be
H_{\mu\nu\rho}=&\nabla_{\mu}R_{\nu\rho}+\nabla_{\nu}R_{\rho\mu}+\nabla_{\rho}R_{\mu\nu}\\
&-\frac{1}{3}\left(g_{\nu\rho}\partial_{\mu}+g_{\rho\mu}\partial_{\nu}+g_{\mu\nu}\partial_{\rho}\right)R~,
\ee
where $R_{\mu\nu}$ is the Ricci tensor, and $R$ is the Ricci scalar. Harada tensor is subjected to the condition
\be
H_{\mu\nu\rho}=8\pi T_{\mu\nu\rho}~,
\ee
where the tensor condition is defined from the energy momentum tensor $T_{\mu\nu}$ and its trace $T$ as
\be
T_{\mu\nu\rho}=&\nabla_{\mu}T_{\nu\rho}+\nabla_{\nu}T_{\rho\mu}+\nabla_{\rho}T_{\mu\nu}\\
&-\frac{1}{6}\left(g_{\nu\rho}\partial_{\mu}+g_{\rho\mu}\partial_{\nu}+g_{\mu\nu}\partial_{\rho}\right)T~.
\ee
\\
It is well established that Schwarzschild metric is obtained from defining $R_{11}$ and $R_{22}$ then plugging them in Einstein's equations. Similarly, one can study $H_{111}$ together with other Ricci tensor components such that eq.\eqref{eq.1} becomes:
\be
ds^2_{\text{H}}=-f(r)dt^2+f(r)^{-1} dr^2+r^2d\theta^2+r^2 \sin^2(\theta)d\phi^2,
\ee
\be \label{eq.4}
f(r)= 1-\frac{2M}{r}-\frac{\Lambda}{3}r^2-\frac{N}{5}r^4~.
\ee
Here $\Lambda$ is the cosmological constant, $M$ is the usual mass, and $N$ is the new Harada parameter that corresponds to implementing the conformal Killing conditions on the energy-momentum tensor. By finding the roots of the \textit{quintic} polynomial \footnote{It is a textbook fact that Galois theory says there is no formula for the roots of quintic equation.}, i.e., $f(r)=0$ or
\be
&3Nr^5+5\Lambda r^3-15r+30M=0~,
\ee
one can define the spatial regions where the inner, event ($r_h$) and cosmological ($r_c$) horizons are located. However, we need to avoid the combinations of $\Lambda$, $M$, and $ N$ that lead to the formation of \textit{extreme} black holes \cite{Sharif:2014ria}, where in such system the event and cosmological horizons coincide. This is crucial for preserving the throat radius $R$ of the wormhole as $r_h < R < r_c$.

\begin{figure}[h!]
\centering
%\begin{minipage}[t]{0.48\linewidth}
\includegraphics[width=\linewidth]{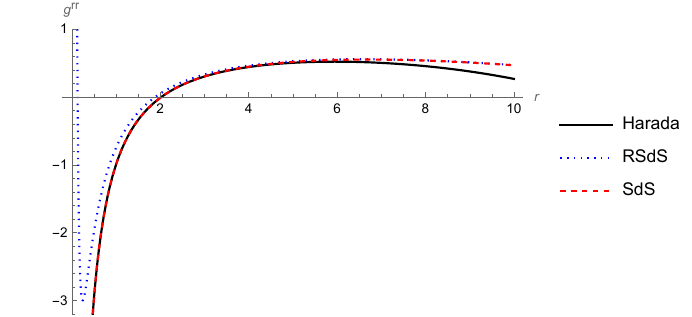}
\caption{The behavior of $g^{rr}$ metric component for (Harada:
$M=1,\Lambda=10^{-2},N=10^{-4}$),\\
Reissner-Nordstrom de-Sitter\\
(RNdS: $M=1,\Lambda=10^{-2},Q=0.5$),\\
and Schwarzschild de-Sitter (SdS: $M=1,\Lambda=10^{-2}$).}\label{fig.1}
\end{figure}

\par\noindent
Now we follow the cut-and-paste technique \cite{Visser:1989kg, Lobo:2003xd} to construct a geodesically complete manifold $\displaystyle{\Gamma=\Gamma_+ \bigcup~\Gamma_-}$. We start by \textit{cutting} spacetime regions $\displaystyle{\Gamma_{\pm} :=\left\lbrace r_{\pm}\leq R~|~R>r_h\right\rbrace}$ inside the throat radius $R$. This is followed by \textit{pasting} the timelike hypersurface regions, named a \textit{thin shell} $\displaystyle{\partial\Gamma=\partial\Gamma_+ \bigcup~\partial\Gamma_-}$~, where $\displaystyle{\partial\Gamma_{\pm} :=\left\lbrace r_{\pm}=R~|~R>r_h\right\rbrace}$, that bounds the bulk of two Harada black holes.\\
\par\noindent
Next, we follow Darmois-Israel-(Sen) junction conditions \cite{Sen:1924,Darmois:1927rt,Israel:1966rt} by defining the $\Gamma$ manifold coordinates as $x^{\mu}:=(t,r,\theta,\phi)$ and $\partial\Gamma$ shell coordinates as $\zeta^i:=(\tau,\theta,\phi)$, where $\tau$ is the proper time measured by a comoving observer when around the throat of the wormhole. The induced metric of the shell is given by
\be
ds^2_{\partial\Gamma}= -d\tau^2 + r^2d\theta^2+r^2 \sin^2(\theta)d\phi^2,
\ee
where the parametric equation $r=R(\tau)$  relates $\Gamma$ to $\partial\Gamma$. So in vacuum spacetime \cite{Lobo:2004id}, the interior solution $r_0$ is matched to the exterior one $R$ at the \emph{junction surface} $\partial\Gamma$ when the surface stress/pressure coefficients are present. We obtain surface stresses using the discontinuity in the extrinsic curvature $\mathcal{K}_{ij}$. Therefore, the thin-shell surface confines the exotic matter to a finite region\footnote{See figure 1 of Ref. \cite{Lobo:2004id}.}. To minimize the violation of the average null energy condition (ANEC), we design the wormhole such that the exotic matter is impounded to the junction region $r_0<r<R$ with the limit $r_0\to R$ that causes the junction to evolve into a thin-shell.\\
\par\noindent
After that, We decompose Harada spacetime using the Gauss-Codazzi approach, and consequently we obtain Israel's junction condition on $\Gamma$ \cite{2740310}. This condition is described by the $3D$ energy-momentum tensor on the junction $\mathcal{S}^i_{~j}=\text{diag}\left(-\sigma,p_{\theta},p_{\phi}\right)$ as
\be
\mathcal{S}_{ij}=-\frac{1}{8\pi}\left(\left[\mathcal{K}_{ij}\right] -\delta_{ij}\mathcal{K}\right),
\ee  
where $\left[\mathcal{K}_{ij}\right]=\mathcal{K}^{+}_{ij}-\mathcal{K}^{-}_{ij}$ is the discontinuity in the extrinsic curvature, and $\mathcal{K}=\left[\mathcal{K}^i_{~i}\right]$ is its trace.\\
We now define the unit vectors $n^{\pm}_\mu$ normal to $\partial\Gamma$ as
\be \label{eq.8}
n^{\pm}_\mu=\pm\left(~\biggl\lvert g^{\alpha\beta}\frac{\partial f}{\partial x^{\alpha}}\frac{\partial f}{\partial x^{\beta}}\biggr\rvert^{-1/2}\frac{\partial f}{\partial x^{\mu}}\right)~.
\ee
Then, the components of the extrinsic curvature, or the second fundamental form, are defined as
\be\label{eq.9}
\mathcal{K}^\pm_{ij}=-n_{\mu}\left(\frac{\partial^2 x^\mu}{\partial\zeta^i\zeta^j}+\Gamma^{\mu\pm}_{\nu\rho}~\frac{\partial x^\nu}{\partial \zeta^i}\frac{\partial x^\rho}{\partial \zeta^j}\right)
\ee
We substitute eq.\eqref{eq.4} in eq.\eqref{eq.8} to get
\be \label{eq.10}
n^{\pm}_\mu=\biggl(\mp \dot{R},\pm\frac{\sqrt{\dot{R}^2+f(R)}}{f(R)},0,0\biggr)~.
\ee
After that, we substitute eq.\eqref{eq.10} in eq.\eqref{eq.9} to get the extrinsic curvature components in Harada gravity as
\be 
&\mathcal{K}^{\pm}_{\theta\theta}=\mathcal{K}^{\pm}_{\phi\phi} =\pm\frac{1}{R}\sqrt{1-\frac{2M}{R}-\frac{\Lambda}{3}R^2-\frac{N}{5}R^4 + \dot{R}^2}~,\\
&\mathcal{K}^{\pm}_{\tau\tau}=\mp\frac{1}{2}\frac{\frac{2M}{R^2}-\frac{2\Lambda}{3}R-\frac{4N}{5}R^3+2\ddot{R}}{\sqrt{1-\frac{2M}{R}-\frac{\Lambda}{3}R^2-\frac{N}{5}R^4 + \dot{R}^2}}~.
\ee
Thus we use these components to define the surface stress and pressure as
\be\label{eq.12}
\sigma &=-\frac{1}{2\pi}\mathcal{K}_{\theta\theta}\\
&=-\frac{1}{2\pi R}\sqrt{1-\frac{2M}{R}-\frac{\Lambda}{3}R^2-\frac{N}{5}R^4 + \dot{R}^2}~,
\ee
\be
p&=p_\theta=p_\phi=\frac{1}{4\pi}\left(\mathcal{K}_{\tau\tau}+\mathcal{K}_{\theta\theta}\right)\\
&=\frac{1}{8\pi R}\bigg(\frac{\frac{2M}{R^2}-\frac{2\Lambda}{3}R-\frac{4N}{5}R^3+2\ddot{R}}{\sqrt{1-\frac{2M}{R}-\frac{\Lambda}{3}R^2-\frac{N}{5}R^4 + \dot{R}^2}}R\\
&\qquad\qquad+2\frac{1-\frac{2M}{R}-\frac{\Lambda}{3}R^2-\frac{N}{5}R^4 + \dot{R}^2}{\sqrt{1-\frac{2M}{R}-\frac{\Lambda}{3}R^2-\frac{N}{5}R^4 + \dot{R}^2}}\bigg).
\ee
And for the static configuration, i.e., $\dot{R}=\ddot{R}=0$, the surface stresses become
\be\label{eq.14}
\sigma_0=-\frac{1}{2\pi R_0}\sqrt{1-\frac{2M}{R_0}-\frac{\Lambda}{3}R_0^2-\frac{N}{5}R_0^4}~,
\ee
\be\label{eq.15}
p_0&=\frac{1}{8\pi R_0}\bigg(\frac{\frac{2M}{R_0^2}-\frac{2\Lambda}{3}R_0-\frac{4N}{5}R_0^3}{\sqrt{1-\frac{2M}{R_0}-\frac{\Lambda}{3}R_0^2-\frac{N}{5}R_0^4}}R_0\\
&\qquad\qquad+2\frac{1-\frac{2M}{R_0}-\frac{\Lambda}{3}R_0^2-\frac{N}{5}R_0^4}{\sqrt{1-\frac{2M}{R_0}-\frac{\Lambda}{3}R_0^2-\frac{N}{5}R_0^4}}\bigg)~.
\ee
The last two equations are used to study the violation of different energy conditions; the surface density $\sigma_0<0$ imposes the violation of the weak energy condition (WEC), meanwhile the null energy condition (NEC), $\sigma_0 +p_0>0$, is maintained as long as $\displaystyle{f(R_0)<\frac{M}{R_0}-\frac{\Lambda}{3}R_0^2-\frac{2N}{5}R_0^4}$ with no additional exotic effects. And for the strong energy condition (SEC), $\sigma_0 +3p_0>0$, it is also maintained as long as $\displaystyle{f(R_0)>\Lambda R_0^2+\frac{6N}{5}R_0^4-\frac{3M}{R_0}}$.
%Graphs
\begin{figure}[h!]
%\centering
\begin{minipage}[t]{\linewidth}
\captionsetup{justification=centering}
\includegraphics[width=\linewidth]{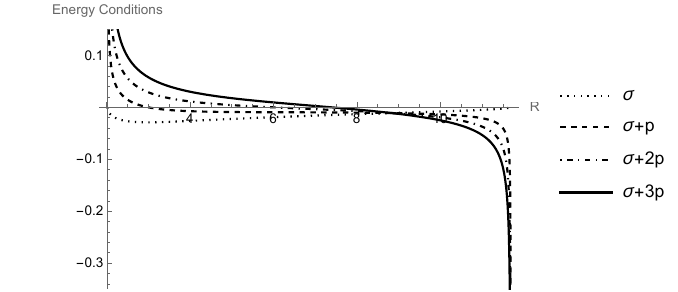}
\subcaption{ $N=10^{-4}$.}
\end{minipage}\hfill

\begin{minipage}[t]{\linewidth}
\captionsetup{justification=centering}
\includegraphics[width=\linewidth]{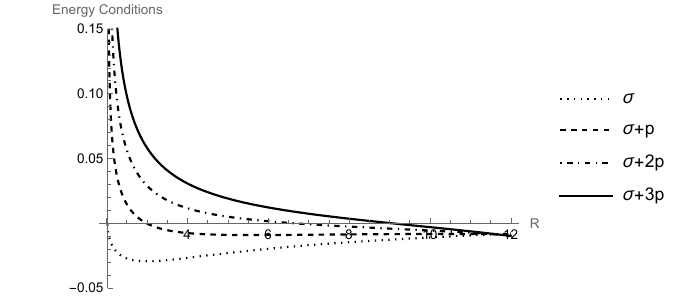}
\subcaption{ $N=10^{-6}$.}
\end{minipage}
\caption{The energy conditions expressed in terms of $\sigma$ and $p$ vs. the throat radius $R_0$ with fixed $M=1$ and $\Lambda=10^{-2}$, and different values of Harada coefficient: $N=10^{-4}$ for figure.2.(a) and $N=10^{-6}$ for figure.2.(b).}\label{fig.2}
\end{figure}
\par\noindent
For Harada black hole with no radial pressure and a mass density localized at the throat $\rho=\sigma_0~\delta(r-R_0)$, the total amount of exotic matter necessary to keep the wormhole open is
\be
\Omega_{\sigma}&=\int^{2\pi}_0\int^{\pi}_0\int^{+\infty}_{-\infty}\sqrt{-g}~\sigma_0~\delta(r-R_0)~dr~d\theta~d\phi\\
&=-2R_0\sqrt{1-\frac{2M}{R_0}-\frac{\Lambda}{3}R_0^2-\frac{N}{5}R_0^4}~.
\ee
Thus we can study the attraction and repulsion characteristics of the corresponding thin-shell wormhole by examining the four-acceleration $a^\mu=u^\nu\nabla_\nu u^\mu$, where the four-velocity $u^\mu=(1/\sqrt{f(r)},0,0,0)$ has only time component $\displaystyle{u^{\mu}\equiv dt/d\tau}$. The geodesic equation of a test particle defines the acceleration as
\be
\frac{d^2 r}{d\tau^2}=-a^r~,
\ee
where in Harada gravity it is given by
\be
a^r=\Gamma^r_{tt}\left(\frac{dt}{d\tau}\right)^2=\frac{6N}{15}r^3+\frac{\Lambda}{3}r-\frac{M}{r^2}~.
\ee
We notice that the wormhole has attractive or repulsive nature if $a^r>0$ or $a^r<0$ respectively.

%Graphs
\begin{figure}[h!]
%\centering
\begin{minipage}[t]{\linewidth}
\captionsetup{justification=centering}
\includegraphics[width=\linewidth]{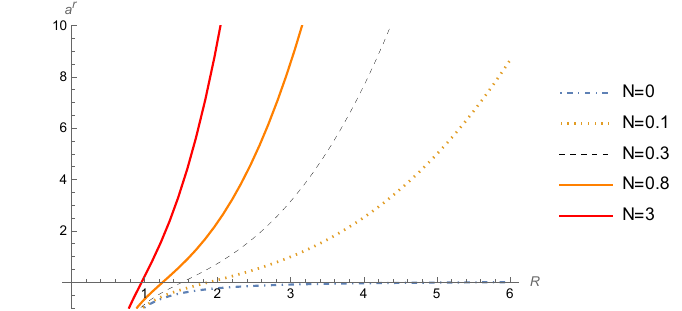}
\subcaption{Different attractive and repulsive behaviors of $a^r$ at small $R$ values for different $N$ values.}
\end{minipage}\hfill

\begin{minipage}[t]{\linewidth}
\captionsetup{justification=centering}
\includegraphics[width=\linewidth]{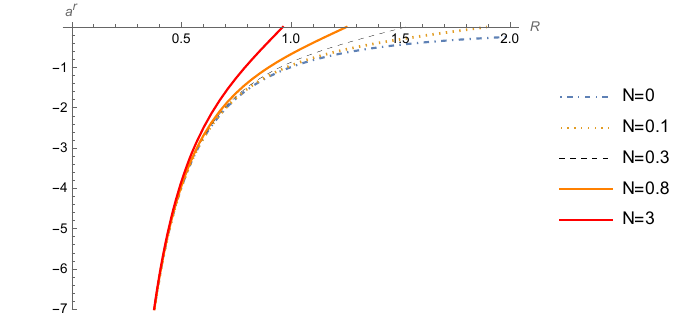}
\subcaption{Convergent behavior of $a^r$ at small $R$ values for different $N$ values. $a^r$ becomes repulsive at almost the same value of $R$ regardless the value of $N$.}
\end{minipage}
\caption{Attraction and repulsion in terms of acceleration $a^r$ vs. the throat radius $R$ with fixed $M=1$ and $\Lambda=10^{-2}$, and different values of Harada coefficient $N$.}\label{fig.3}
\end{figure}
%
%
%
%
%%%%%%%%%%%%%%
\section{Linearized Stability Analysis}
\par\noindent
We check the stability of the wormhole by performing linear perturbation for eq.\eqref{eq.14} and eq.\eqref{eq.15} around the static configuration \cite{Lobo:2003xd} ,i.e., when $(R=R_0)$. Then, we differentiate eq.\eqref{eq.12} with respect to $\tau$ to obtain the continuity equation
\be
\frac{d(\sigma A)}{d\tau}+p\frac{dA}{d\tau}=0~,
\ee
which directly yields
\be \label{eq.20}
\sigma'=-\frac{2}{R}(\sigma+p)~,
\ee
where $A=4\pi R^2$ is the area of the wormhole throat, and $\sigma'=\dot{\sigma}/\dot{R}$; the dot is for $d/d\tau$, and the prime is for $d/dR$.\\
Next, rearrange eq.\eqref{eq.12} and define a potential function
\be \label{eq.21}
V(R)=f(R)-4\pi^2 R^2 \sigma^2=-\dot{R}~.
\ee
Then, we substitute eq.\eqref{eq.20} in the first derivative of eq.\eqref{eq.21} to find that
\be\label{eq.22}
V'(R)= \frac{2M}{R^2}-\frac{2\Lambda}{3}R-\frac{4N}{5}R^3+8\pi^2 R\sigma(\sigma+2p)~.
\ee
And for the second derivative of \eqref{eq.21}, pressure is parameterized to be a function in the density $p:=p(\sigma)$ \cite{Poisson:1995sv}. Thus we introduce another parameter $\vartheta(\sigma)=dp/d\sigma$, which can be seen as the ``speed of sound''. Therefore, the second derivative of \eqref{eq.21} becomes
\be\label{eq.23}
V''(R)&=f''(R)-8\pi^2\left[2\sigma(\sigma+p)(1+2\vartheta)+(\sigma+2p)^2\right]\\
&=f''(R)+\Bigg[~~\frac{1}{R^2}\bigg(Rf'(R)-2f(R)\bigg)\bigg(1+2\vartheta\bigg)\\
&\qquad\qquad\quad~~-\frac{1}{2}\frac{\left(f'(R)\right)^2}{f(R)}\Bigg]~.
\ee
In order to linearize the time rate change of the radius $\dot{R}$, we utilize Taylor expansion with the potential function around the static point $R=R_0$
\be
V(R)=&V(R_0)+(R-R_0)V'(R_0)\\
&+\frac{1}{2}(R-R_0)^2V''(R_0)+\mathcal{O}\left[(R-R_0)^3\right]~.
\ee
We use eq.\eqref{eq.14} and eq.\eqref{eq.15} to evaluate eq.\eqref{eq.21} and eq.\eqref{eq.22} at $R=R_0$. So we notice that the first two terms in the expansion vanish $V(R_0)=V'(R_0)=0$. Meanwhile the second derivative term eq.\eqref{eq.23} becomes
\be\label{eq.25}
V''(R_0)=&-\frac{4M}{R_0^3}-\frac{2\Lambda}{3}-\frac{12N}{5}R_0^2\\
&+\bigg[\frac{2M}{R_0^3}-\frac{2\Lambda}{3}-\frac{4N}{5}R_0^2\\
&-\frac{2}{R_0}+\frac{4M}{R_0^2}+\frac{2\Lambda}{3}R_0+\frac{2N}{5}R_0^3\bigg]\bigg[1+2\vartheta_0\bigg]\\
&-\frac{1}{2}\frac{\left(\frac{2M}{R_0^2}-\frac{2\Lambda}{3}R_0-\frac{4N}{5}R_0^3\right)^2}{1-\frac{2M}{R_0}-\frac{\Lambda}{3}R^2_0-\frac{N}{5}R_0^4}~.
\ee
On can use $(1+2\vartheta)=(\sigma'+2p')/\sigma'$ to express $\vartheta$ in terms of the metric parameters $M,~N$ and $\Lambda$ such that $V''$ has no implicit dependency of the metric parameters $\vartheta(\Lambda,M,N)$. But our current focus is to study the behavior of $\vartheta$ when the throat is stable.\\
%\par\noindent
The concave down test $V''(R_0)<0$ results in causing either expansion or contraction of the throat when small perturbation takes place. While the convex, or the concave up, condition $V''(R_0)>0$ stabilizes the throat with a local minimum of $V(R_0)$. Therefore, we solve eq.\eqref{eq.25} for $\vartheta_0$ at that local minimum to get
\be\label{eq.28}
\vartheta_0 <\frac{1}{2}\bigg[\frac{\frac{\left(\frac{2M}{R_0^2}-\frac{2\Lambda}{3}R_0-\frac{4N}{5}R_0^3\right)^2}{2(1-\frac{2M}{R_0}-\frac{\Lambda}{3}R^2_0-\frac{N}{5}R_0^4)}+\frac{4M}{R_0^3}+\frac{2\Lambda}{3}+\frac{12N}{5}R_0^2}{\frac{2M}{R_0^3}-\frac{2\Lambda}{3}-\frac{4N}{5}R_0^2-\frac{2}{R_0}+\frac{4M}{R_0^2}+\frac{2\Lambda}{3}R_0+\frac{2N}{5}R_0^3}-1\bigg]~.
\ee
\begin{widetext}
\onecolumngrid
%\vspace*{-5cm}
%Graphs
\begin{figure}[!h]
%\centering
\begin{minipage}[t]{0.48\linewidth}
\captionsetup{justification=centering}
\includegraphics[width=\linewidth]{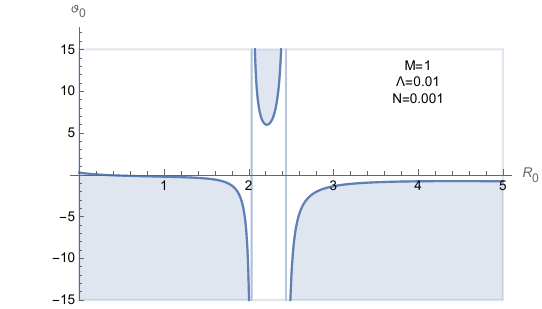}
\subcaption{ $N=0.001$.}
\end{minipage}\hfill
\begin{minipage}[t]{0.48\linewidth}
\captionsetup{justification=centering}
\includegraphics[width=\linewidth]{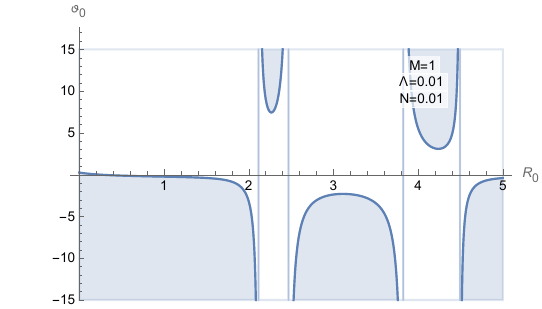}
\subcaption{ $N=0.01$.}
\end{minipage}

\begin{minipage}[t]{0.48\linewidth}
\captionsetup{justification=centering}
\includegraphics[width=\linewidth]{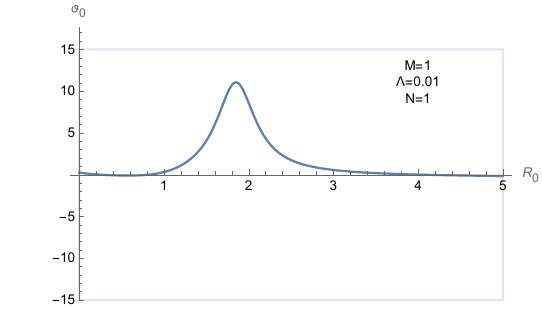}
\subcaption{ $N=1$.}
  \end{minipage}\hfill
\begin{minipage}[t]{0.48\linewidth}
\captionsetup{justification=centering}
\includegraphics[width=\linewidth]{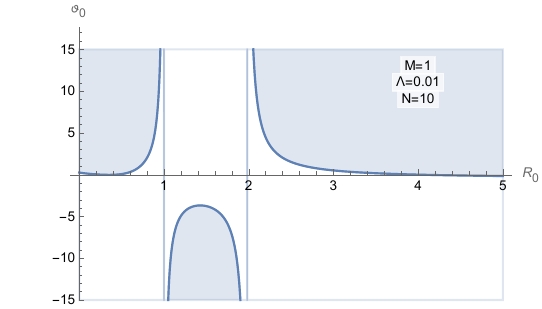}
\subcaption{ $N=10$.}
  \end{minipage}

\caption{Regions of stability of the thin-shell wormhole for the Harada solution with fixed values of $M=1$ and $\Lambda=10^{-2}$, and different values of $N$. Stable regions are the blue shaded domains.}\label{fig.4}
\end{figure}

\end{widetext}
\pagebreak
\twocolumngrid

%%%%%%%%%%%%%%
%
%
%%%%%%%%%%%%%%

\section{Discussion}
\par\noindent
In this letter we design Harada thin-shell wormholes. We utilize Visser's technique of cut-and-paste with Darmois-Israel-(Sen) junction condition to connect two Harada regions of spacetime through a thin shell. We compare the asymptotic behavior of Harada metric with that of SdS and RNdS as in fig.(\ref{fig.1}). Also, We examine the components of the energy-momentum hypersurface tensor using the second fundamental form. Next, we find that WEC is always violated. However, both NEC and SEC can be maintained upon imposing the inequalities that relate $f(r)$ to $f'(r)$. The energy conditions are shown in fig.(\ref{fig.2}). Then, we study the radial acceleration to examine the attractive and repulsive nature of the wormhole throat. The results are plotted in fig.(\ref{fig.3}).\\
%\par\noindent
After that, we analyze the linearized stability of Harada thin-shell wormhole by examining the concavity behavior of eq.\eqref{eq.28}, that of the ``speed of sound'', as a function in Harada parameter, the mass, and the cosmological constant. And we notice changes in stability regions upon varying Harada parameter while keeping the mass and cosmological constant fixed. Besides being considered as a stability measure for the wormhole solutions, the speed of sound function in principle can help detecting such systems by revealing the thermodynamics of the equations of states of the ordinary and/or exotic matter supporting the wormhole system. The analysis of stability regions of speed of sound is demonstrated in fig.(\ref{fig.4}), and we conclude that for a lesser value of cosmological constant and lesser/larger value of magnetic charge, relative to the amount of mass, regions of stability vary. Once the mass is equal to the Harada parameter, we no longer have stability regions. Therefore in the Harada spacetime, and for a small value of the cosmological constant, thin-shell wormholes are kept open as long as the value of Harada parameter is away from the value of mass.
%%%%%%%%%%%%%%
\begin{widetext}
%\onecolumngrid
\section{Acknowledgement}
The authors would like to thank the Department of Physics, Santa Clara University, especially the Chair Prof. Dr. G. Ramon, for supporting the senior undergraduate physics seminars where the authors met and worked to produce this letter.
\end{widetext}

%%%%%%%%%%%%%%

%%%%%%%%%%%%%%
%
%

\end{arabicfootnotes}

\begin{thebibliography}{9}

%\cite{Morris:1988cz}
\bibitem{Morris:1988cz} 
  M.~S.~Morris and K.~S.~Thorne,
  %``Wormholes in space-time and their use for interstellar travel: A tool for teaching general relativity,''
  Am.\ J.\ Phys.\  {\bf 56}, 395 (1988).

%\cite{Visser:1995cc}
\bibitem{Visser:1995cc} 
  M.~Visser,
  ``Lorentzian wormholes: From Einstein to Hawking,''
  Woodbury, USA: AIP (1995) 412 p.

%\cite{Lobo:2005vc}
\bibitem{Lobo:2005vc} 
  F.~S.~N.~Lobo,
  %``Chaplygin traversable wormholes,''
  Phys.\ Rev.\ D {\bf 73}, 064028 (2006)
  [gr-qc/0511003].

%\cite{Sushkov:2005kj}
\bibitem{Sushkov:2005kj} 
  S.~V.~Sushkov,
  %``Wormholes supported by a phantom energy,''
  Phys.\ Rev.\ D {\bf 71}, 043520 (2005)
  [gr-qc/0502084].

%\cite{Lobo:2005yv}
\bibitem{Lobo:2005yv} 
  F.~S.~N.~Lobo,
  %``Stability of phantom wormholes,''
  Phys.\ Rev.\ D {\bf 71}, 124022 (2005)
  [gr-qc/0506001].

%\cite{Lobo:2006ue}
\bibitem{Lobo:2006ue} 
  F.~S.~N.~Lobo,
  %``Van der Waals quintessence stars,''
  Phys.\ Rev.\ D {\bf 75}, 024023 (2007)
  [gr-qc/0610118].

%\cite{Poisson:1995sv}
\bibitem{Poisson:1995sv} 
  E.~Poisson and M.~Visser,
  %``Thin shell wormholes: Linearization stability,''
  Phys.\ Rev.\ D {\bf 52}, 7318 (1995)
  [gr-qc/9506083].

%\cite{Kar:1995ss}
\bibitem{Kar:1995ss} 
  S.~Kar and D.~Sahdev,
  %``Evolving Lorentzian wormholes,''
  Phys.\ Rev.\ D {\bf 53}, 722 (1996)
  [gr-qc/9506094].

%\cite{Teo:1998dp}
\bibitem{Teo:1998dp} 
  E.~Teo,
  %``Rotating traversable wormholes,''
  Phys.\ Rev.\ D {\bf 58}, 024014 (1998)
  [gr-qc/9803098].

%\cite{Lobo:2004rp}
\bibitem{Lobo:2004rp} 
  F.~S.~N.~Lobo,
  %``Energy conditions, traversable wormholes and dust shells,''
  Gen.\ Rel.\ Grav.\  {\bf 37}, 2023 (2005)
  [gr-qc/0410087].

%\cite{Garattini:2019ivd}
\bibitem{Garattini:2019ivd} 
  R.~Garattini,
  %``Casimir Wormholes,''
  Eur.\ Phys.\ J.\ C {\bf 79}, no. 11, 951 (2019)
  [arXiv:1907.03623 [gr-qc]].
  
%\cite{Kuhfittig:2010pb}
\bibitem{Kuhfittig:2010pb} 
  P.~K.~F.~Kuhfittig,
  %``The Stability of thin-shell wormholes with a phantom-like equation of state,''
  Acta Phys.\ Polon.\ B {\bf 41}, 2017 (2010)
  [arXiv:1008.3111 [gr-qc]].
  
%\cite{Rahaman:2011yh}
\bibitem{Rahaman:2011yh} 
  F.~Rahaman, A.~Banerjee and I.~Radinschi,
  %``A new class of stable $(2+1)$ dimensional thin shell wormhole,''
  Int.\ J.\ Theor.\ Phys.\  {\bf 51}, 1680 (2012)
  [arXiv:1109.0976 [gr-qc]].    
  
%\cite{Sharif:2013xta}
\bibitem{Sharif:2013xta} 
  M.~Sharif and M.~Azam,
  %``Stability Analysis of Thin-Shell Wormholes from Charged Black String,''
  JCAP {\bf 1304}, 023 (2013)
  [arXiv:1305.4441 [gr-qc]].  
  
%\cite{Sharif:2014ria}
\bibitem{Sharif:2014ria} 
  M.~Sharif and S.~Mumtaz,
  %``Schwarzschild-de Sitter and Anti-de Sitter Thin-Shell Wormholes and Their Stability,''
  Adv.\ High Energy Phys.\  {\bf 2014}, 639759 (2014).
  
%\cite{Eid:2015pja}
\bibitem{Eid:2015pja} 
  A.~Eid,
  %``Linearized stability of Reissner Nordstrom de-Sitter thin shell wormholes,''
  New Astron.\  {\bf 39}, 72 (2015).  

%\cite{Ovgun:2017jzt}
\bibitem{Ovgun:2017jzt} 
  A.~Övgün, A.~Banerjee and K.~Jusufi,
  %``Charged Thin-shell Gravastars in Noncommutative Geometry,''
  Eur.\ Phys.\ J.\ C {\bf 77}, no. 8, 566 (2017)
  [arXiv:1704.00603 [gr-qc]].    
  
%\cite{Ishak:2001az}
\bibitem{Ishak:2001az} 
  M.~Ishak and K.~Lake,
  %``Stability of transparent spherically symmetric thin shells and wormholes,''
  Phys.\ Rev.\ D {\bf 65}, 044011 (2002)
  [gr-qc/0108058].  

%\cite{Lobo:2005zu}
\bibitem{Lobo:2005zu} 
  F.~S.~N.~Lobo and P.~Crawford,
  %``Stability analysis of dynamic thin shells,''
  Class.\ Quant.\ Grav.\  {\bf 22}, 4869 (2005)
  [gr-qc/0507063].

%\cite{Eiroa:2007qz}
\bibitem{Eiroa:2007qz} 
  E.~F.~Eiroa and C.~Simeone,
  %``Stability of Chaplygin gas thin-shell wormholes,''
  Phys.\ Rev.\ D {\bf 76}, 024021 (2007)
  [arXiv:0704.1136 [gr-qc]].

%\cite{Eiroa:2008ky}
\bibitem{Eiroa:2008ky} 
  E.~F.~Eiroa,
  %``Stability of thin-shell wormholes with spherical symmetry,''
  Phys.\ Rev.\ D {\bf 78}, 024018 (2008)
  [arXiv:0805.1403 [gr-qc]].

%\cite{Lemos:2008aj}
\bibitem{Lemos:2008aj} 
  J.~P.~S.~Lemos and F.~S.~N.~Lobo,
  %``Plane symmetric thin-shell wormholes: Solutions and stability,''
  Phys.\ Rev.\ D {\bf 78}, 044030 (2008)
  [arXiv:0806.4459 [gr-qc]].

%\cite{Dias:2010uh}
\bibitem{Dias:2010uh} 
  G.~A.~S.~Dias and J.~P.~S.~Lemos,
  %``Thin-shell wormholes in $d$-dimensional general relativity: Solutions, properties, and stability,''
  Phys.\ Rev.\ D {\bf 82}, 084023 (2010)
  [arXiv:1008.3376 [gr-qc]].

%\cite{Eiroa:2011nd}
\bibitem{Eiroa:2011nd} 
  E.~F.~Eiroa and C.~Simeone,
  %``Stability of charged thin shells,''
  Phys.\ Rev.\ D {\bf 83}, 104009 (2011)
  [arXiv:1102.1683 [gr-qc]].

%\cite{Sharif:2013nka}
\bibitem{Sharif:2013nka} 
  M.~Sharif and M.~Azam,
  %``Stability of Thin-Shell Wormholes in Nonlinear Electrodynamics,''
  J.\ Phys.\ Soc.\ Jap.\  {\bf 81}, 124006 (2012)
  [arXiv:1307.1100 [gr-qc]].

%\cite{Mazharimousavi:2014gpa}
\bibitem{Mazharimousavi:2014gpa} 
  S.~H.~Mazharimousavi, M.~Halilsoy and Z.~Amirabi,
  %``Stability of generic cylindrical thin shell wormholes,''
  Phys.\ Rev.\ D {\bf 89}, no. 8, 084003 (2014)
  [arXiv:1403.2861 [gr-qc]].

%\cite{Lobo:2015lbc}
\bibitem{Lobo:2015lbc} 
  F.~S.~N.~Lobo, P.~Martín-Moruno, N.~Montelongo-García and M.~Visser,
  %``Novel stability approach of thin-shell gravastars,''
  arXiv:1512.07659 [gr-qc].

%\cite{Eid:2016axb}
\bibitem{Eid:2016axb} 
  A.~Eid,
  %``On the stability of charged thin-shell wormholes,''
  Eur.\ Phys.\ J.\ Plus {\bf 131}, no. 2, 23 (2016).

%\cite{Ovgun:2017jip}
\bibitem{Ovgun:2017jip} 
  A.~Ovgün and K.~Jusufi,
  %``Stability of Effective Thin-shell Wormholes Under Lorentz Symmetry Breaking Supported by Dark Matter and Dark Energy,''
  Eur.\ Phys.\ J.\ Plus {\bf 132}, no. 12, 543 (2017)
  [arXiv:1706.07656 [gr-qc]].

%\cite{Amirabi:2017buh}
\bibitem{Amirabi:2017buh} 
  Z.~Amirabi,
  %``Stability of generic thin shells in conformally flat spacetimes,''
  Eur.\ Phys.\ J.\ C {\bf 77}, no. 7, 493 (2017).
  
%\cite{HabibMazharimousavi:2017zlc}
\bibitem{HabibMazharimousavi:2017zlc} 
  S.~Habib Mazharimousavi, M.~Halilsoy and S.~N.~Hamad Amen,
  %``Stability of spherically symmetric timelike thin-shells in general relativity with a variable equation of state,''
  Int.\ J.\ Mod.\ Phys.\ D {\bf 26}, no. 14, 1750158 (2017)
  [arXiv:1708.04588 [gr-qc]].  

%\cite{Eiroa:2017nar}
\bibitem{Eiroa:2017nar} 
  E.~F.~Eiroa and G.~Figueroa Aguirre,
  %``Spherical thin shells in $F(R)$ gravity: construction and stability,''
  Eur.\ Phys.\ J.\ C {\bf 78}, no. 1, 54 (2018)
  [arXiv:1711.02583 [gr-qc]].

%\cite{Tsukamoto:2018lsg}
\bibitem{Tsukamoto:2018lsg} 
  N.~Tsukamoto and T.~Kokubu,
  %``Linear stability analysis of a rotating thin-shell wormhole,''
  Phys.\ Rev.\ D {\bf 98}, no. 4, 044026 (2018)
  [arXiv:1807.01528 [gr-qc]].

%\cite{Forghani:2019wgt}
\bibitem{Forghani:2019wgt} 
  S.~D.~Forghani, S.~H.~Mazharimousavi and M.~Halilsoy,
  %``Discontinuity Problem in the Linear Stability Analysis of Thin-Shell Wormholes,''
  Eur.\ Phys.\ J.\ Plus {\bf 134}, no. 7, 342 (2019)
  [arXiv:1903.02035 [gr-qc]].
  
%\cite{Halilsoy:2013iza}
\bibitem{Halilsoy:2013iza} 
  M.~Halilsoy, A.~Ovgun and S.~H.~Mazharimousavi,
  %``Thin-shell wormholes from the regular Hayward black hole,''
  Eur.\ Phys.\ J.\ C {\bf 74}, 2796 (2014)
  [arXiv:1312.6665 [gr-qc]].  
  
%\cite{Sharif:2016gyb}
\bibitem{Sharif:2016gyb} 
  M.~Sharif and S.~Mumtaz,
  %``Stability of thin-shell wormholes from a regular ABG black hole,''
  Eur.\ Phys.\ J.\ Plus {\bf 132}, no. 1, 26 (2017)
  [arXiv:1604.01012 [gr-qc]].

%\cite{Alshal:2019owh}
\bibitem{Alshal:2019owh}
H.~Alshal,
%``Linearized Stability of Bardeen de-Sitter Thin-Shell Wormholes,''
EPL \textbf{128}, no.6, 60007 (2019)
[arXiv:1909.07811 [gr-qc]].

%\cite{Konoplya:2021hsm}
\bibitem{Konoplya:2021hsm}
R.~A.~Konoplya and A.~Zhidenko,
%``Traversable Wormholes in General Relativity,''
Phys. Rev. Lett. \textbf{128} (2022) no.9, 091104
doi:10.1103/PhysRevLett.128.091104
[arXiv:2106.05034 [gr-qc]].

%\cite{Churilova:2021tgn}
\bibitem{Churilova:2021tgn}
M.~S.~Churilova, R.~A.~Konoplya, Z.~Stuchlik and A.~Zhidenko,
%``Wormholes without exotic matter: quasinormal modes, echoes and shadows,''
JCAP \textbf{10} (2021), 010
doi:10.1088/1475-7516/2021/10/010
[arXiv:2107.05977 [gr-qc]].

%\cite{DiGrezia:2017daq}
\bibitem{DiGrezia:2017daq}
E.~Di Grezia, E.~Battista, M.~Manfredonia and G.~Miele,
%``Spin, torsion and violation of null energy condition in traversable wormholes,''
Eur. Phys. J. Plus \textbf{132} (2017) no.12, 537
doi:10.1140/epjp/i2017-11799-6
[arXiv:1707.01508 [gr-qc]].

%\cite{Bronnikov:2018uje}
\bibitem{Bronnikov:2018uje}
K.~A.~Bronnikov and V.~G.~Krechet,
%``Potentially observable cylindrical wormholes without exotic matter in general relativity,''
Phys. Rev. D \textbf{99} (2019) no.8, 084051
doi:10.1103/PhysRevD.99.084051
[arXiv:1807.03641 [gr-qc]].

%\cite{Kuhfittig:2020zmp}
\bibitem{Kuhfittig:2020zmp}
P.~K.~F.~Kuhfittig,
%``Noncommutative-geometry wormholes without exotic matter,''
Adv. Stud. Theor. Phys. \textbf{14} (2020) no.5-8, 219-225
[arXiv:2008.06728 [gr-qc]].

%\cite{DeFalco:2021ksd}
\bibitem{DeFalco:2021ksd}
V.~De Falco, E.~Battista, S.~Capozziello and M.~De Laurentis,
%``Reconstructing wormhole solutions in curvature based Extended Theories of Gravity,''
Eur. Phys. J. C \textbf{81} (2021) no.2, 157
doi:10.1140/epjc/s10052-021-08958-4
[arXiv:2102.01123 [gr-qc]].

%\cite{Klinkhamer:2022rsj}
\bibitem{Klinkhamer:2022rsj}
F.~R.~Klinkhamer,
%``Defect Wormhole: A Traversable Wormhole Without Exotic Matter,''
Acta Phys. Polon. B \textbf{54} (2023) no.5, 5-A3
doi:10.5506/APhysPolB.54.5-A3
[arXiv:2301.00724 [gr-qc]].

%\cite{DeFalco:2023twb}
\bibitem{DeFalco:2023twb}
V.~De Falco and S.~Capozziello,
%``Static and spherically symmetric wormholes in metric-affine theories of gravity,''
Phys. Rev. D \textbf{108} (2023) no.10, 104030
doi:10.1103/PhysRevD.108.104030
[arXiv:2308.05440 [gr-qc]].

%\cite{Gonzalez:2008wd}
\bibitem{Gonzalez:2008wd}
J.~A.~Gonzalez, F.~S.~Guzman and O.~Sarbach,
%``Instability of wormholes supported by a ghost scalar field. I. Linear stability analysis,''
Class. Quant. Grav. \textbf{26} (2009), 015010
doi:10.1088/0264-9381/26/1/015010
[arXiv:0806.0608 [gr-qc]].

%\cite{Bronnikov:2012ch}
\bibitem{Bronnikov:2012ch}
K.~A.~Bronnikov, R.~A.~Konoplya and A.~Zhidenko,
%``Instabilities of wormholes and regular black holes supported by a phantom scalar field,''
Phys. Rev. D \textbf{86} (2012), 024028
doi:10.1103/PhysRevD.86.024028
[arXiv:1205.2224 [gr-qc]].

%\cite{Cuyubamba:2018jdl}
\bibitem{Cuyubamba:2018jdl}
M.~A.~Cuyubamba, R.~A.~Konoplya and A.~Zhidenko,
%``No stable wormholes in Einstein-dilaton-Gauss-Bonnet theory,''
Phys. Rev. D \textbf{98} (2018) no.4, 044040
doi:10.1103/PhysRevD.98.044040
[arXiv:1804.11170 [gr-qc]].

%\cite{Harada:2023rqw}
\bibitem{Harada:2023rqw}
J.~Harada,
%``Gravity at cosmological distances: Explaining the accelerating expansion without dark energy,''
Phys. Rev. D \textbf{108}, no.4, 044031 (2023)
doi:10.1103/PhysRevD.108.044031
[arXiv:2308.02115 [gr-qc]].

%\cite{Harada:2023afu}
\bibitem{Harada:2023afu}
J.~Harada,
%``Dark energy in conformal Killing gravity,''
Phys. Rev. D \textbf{108}, no.10, 104037 (2023)
doi:10.1103/PhysRevD.108.104037
[arXiv:2308.07634 [gr-qc]].

%\cite{Mantica:2023stl}
\bibitem{Mantica:2023stl}
C.~A.~Mantica and L.~G.~Molinari,
%``Note on Harada\textquoteright{}s conformal Killing gravity,''
Phys. Rev. D \textbf{108}, no.12, 124029 (2023)
doi:10.1103/PhysRevD.108.124029
[arXiv:2308.06803 [gr-qc]].

%\cite{Mantica:2024mun}
\bibitem{Mantica:2024mun}
C.~A.~Mantica and L.~G.~Molinari,
%``Conformal Killing cosmology: Geometry, dark sector, growth of structures, and a big rip,''
Phys. Rev. D \textbf{110} (2024) no.6, 064041
doi:10.1103/PhysRevD.110.064041
[arXiv:2404.11468 [gr-qc]].

%\cite{Mantica:2024sdy}
\bibitem{Mantica:2024sdy}
C.~A.~Mantica and L.~G.~Molinari,
%``Conformal Killing gravity in static spherically symmetric spacetimes,''
Phys. Rev. D \textbf{110} (2024) no.4, 044025
doi:10.1103/PhysRevD.110.044025
[arXiv:2406.12511 [gr-qc]].

%\cite{Junior:2023ixh}
\bibitem{Junior:2023ixh}
J.~T.~S.~S.~Junior, F.~S.~N.~Lobo and M.~E.~Rodrigues,
%``(Regular) Black holes in conformal Killing gravity coupled to nonlinear electrodynamics and scalar fields,''
Class. Quant. Grav. \textbf{41}, no.5, 055012 (2024)
doi:10.1088/1361-6382/ad210e
[arXiv:2310.19508 [gr-qc]].

%\cite{Barnes:2023uru}
\bibitem{Barnes:2023uru}
A.~Barnes,
%``Vacuum Static Spherically Symmetric Spacetimes in Harada's Theory,''
[arXiv:2309.05336 [gr-qc]].

%\cite{Clement:2024xmr}
\bibitem{Clement:2024xmr}
G.~Cl\'ement and K.~Nouicer,
%``Spherical symmetric solutions of conformal Killing gravity: black holes, wormholes, and sourceless cosmologies,''
Class. Quant. Grav. \textbf{41}, no.16, 165005 (2024)
doi:10.1088/1361-6382/ad5f24
[arXiv:2404.00328 [gr-qc]].

%\cite{Junior:2024vrv}
\bibitem{Junior:2024vrv}
J.~T.~S.~S.~Junior, F.~S.~N.~Lobo and M.~E.~Rodrigues,
%``Black bounces in conformal Killing gravity,''
Eur. Phys. J. C \textbf{84}, no.6, 557 (2024)
doi:10.1140/epjc/s10052-024-12922-3
[arXiv:2405.09702 [gr-qc]].

%\cite{Barnes:2024vjq}
\bibitem{Barnes:2024vjq}
A.~Barnes,
%``pp-waves in conformal Killing gravity,''
[arXiv:2404.09310 [gr-qc]].

%\cite{Morris:1988tu}
\bibitem{Morris:1988tu}
M.~S.~Morris, K.~S.~Thorne and U.~Yurtsever,
%``Wormholes, Time Machines, and the Weak Energy Condition,''
Phys. Rev. Lett. \textbf{61}, 1446-1449 (1988)
doi:10.1103/PhysRevLett.61.1446

%\cite{Visser:1989kg}
\bibitem{Visser:1989kg} 
  M.~Visser,
  %``Traversable wormholes from surgically modified Schwarzschild space-times,''
  Nucl.\ Phys.\ B {\bf 328}, 203 (1989)
  [arXiv:0809.0927 [gr-qc]].

%\cite{Visser:1989kh}
\bibitem{Visser:1989kh} 
  M.~Visser,
  %``Traversable wormholes: Some simple examples,''
  Phys.\ Rev.\ D {\bf 39}, 3182 (1989)
  [arXiv:0809.0907 [gr-qc]].

%\cite{Sen:1924}
\bibitem{Sen:1924} 
  N.~Sen,
  %``Über die Grenzbedingungen des Schwerefeldes an Unstetigkeitsflächen,''
  Ann.\ Phys.\, {\bf 378}: 365-396. (1924).

%\cite{Israel:1966rt}
\bibitem{Israel:1966rt} 
  W.~Israel,
  %``Singular hypersurfaces and thin shells in general relativity,''
  Nuovo Cim.\ B {\bf 44S10}, 1 (1966)
  [Nuovo Cim.\ B {\bf 44}, 1 (1966)]
  Erratum: [Nuovo Cim.\ B {\bf 48}, 463 (1967)].

\bibitem{Darmois:1927rt}
G. Darmois, Memorial de Sciences Mathematiques, Fascicule XXV, ”Les equations de la gravitation einsteinienne”, Chapitre V (1927).

%\cite{Mansouri:1996ps}
\bibitem{Mansouri:1996ps} 
  R.~Mansouri and M.~Khorrami,
  %``Equivalence of Darmois-Israel and distributional methods for thin shells in general relativity,''
  J.\ Math.\ Phys.\  {\bf 37}, 5672 (1996)
  [gr-qc/9608029].

%\cite{2740310}
\bibitem{2740310}
S.~Khakshournia and R.~Mansouri,
``The Art of Gluing Space-Time Manifolds: Methods and Applications'', 2023, Springer Cham, ISBN: 9783031486128
%doi:10.1007/978-3-031-48612-8

%\cite{Sharif:2013tva}
\bibitem{Sharif:2013tva}
M.~Sharif and M.~Azam,
%``Spherical thin-shell wormholes and modified Chaplygin gas,''
JCAP \textbf{05}, 025 (2013)
doi:10.1088/1475-7516/2013/05/025
[arXiv:1310.0326 [gr-qc]].

%\cite{Godani:2022hik}
\bibitem{Godani:2022hik}
N.~Godani,
%``Linear stability of Bardeen anti-de Sitter thin-shell wormhole in f(R) gravity,''
Int. J. Geom. Meth. Mod. Phys. \textbf{19}, no.13, 2250208 (2022)

%\cite{Eid:2023fbh}
\bibitem{Eid:2023fbh}
A.~Eid,
%``Dynamics and stability of Bardeen-de Sitter thin shell wormholes,''
New Astron. \textbf{98}, 101934 (2023)

%\cite{Eid:2023wrd}
\bibitem{Eid:2023wrd}
A.~Eid and A.~Alkaoud,
%``Dynamics and stability of Hayward -de Sitter thin-shell wormhole,''
New Astron. \textbf{101}, 102021 (2023)

%\cite{Sharif:2021ugf}
\bibitem{Sharif:2021ugf}
M.~Sharif and F.~Javed,
%``Stability of Charged Thin-Shell Wormholes with Weyl Corrections,''
Astron. Rep. \textbf{65}, no.5, 353-361 (2021)

%\cite{Sharif:2021zzr}
\bibitem{Sharif:2021zzr}
M.~Sharif and F.~Javed,
%``Stability and Dynamics of Regular Thin-Shell Gravastars,''
J. Exp. Theor. Phys. \textbf{132}, no.3, 381-393 (2021)

%\cite{Sharif:2021cvr}
\bibitem{Sharif:2021cvr}
M.~Sharif and F.~Javed,
%``Stability of charged thin-shell and thin-shell wormholes: a comparison,''
Phys. Scripta \textbf{96}, no.5, 055003 (2021)

%\cite{Sharif:2020nhc}
\bibitem{Sharif:2020nhc}
M.~Sharif and F.~Javed,
%``Stability of charged Kiselev thin-shell wormholes,''
Int. J. Mod. Phys. A \textbf{35}, no.02n03, 2040015 (2020)

%\cite{Eiroa:2003wp}
\bibitem{Eiroa:2003wp}
E.~F.~Eiroa and G.~E.~Romero,
%``Linearized stability of charged thin shell wormholes,''
Gen. Rel. Grav. \textbf{36}, 651-659 (2004)
doi:10.1023/B:GERG.0000016916.79221.24
[arXiv:gr-qc/0303093 [gr-qc]].

%\cite{Bejarano:2016gyv}
\bibitem{Bejarano:2016gyv}
C.~Bejarano, F.~S.~N.~Lobo, G.~J.~Olmo and D.~Rubiera-Garcia,
%``Palatini wormholes and energy conditions from the prism of General Relativity,''
Eur. Phys. J. C \textbf{77}, no.11, 776 (2017)
doi:10.1140/epjc/s10052-017-5353-0
[arXiv:1607.01259 [gr-qc]].

%\cite{Lobo:2003xd}
\bibitem{Lobo:2003xd} 
  F.~S.~N.~Lobo and P.~Crawford,
  %``Linearized stability analysis of thin shell wormholes with a cosmological constant,''
  Class.\ Quant.\ Grav.\  {\bf 21}, 391 (2004)
  [gr-qc/0311002].
 
%\cite{Lobo:2004id}
\bibitem{Lobo:2004id} 
  F.~S.~N.~Lobo,
  %``Surface stresses on a thin shell surrounding a traversable wormhole,''
  Class.\ Quant.\ Grav.\  {\bf 21}, 4811 (2004)
  [gr-qc/0409018].  
  
\end{thebibliography}
\end{document}